\documentclass[aps,pr2,twocolumn,superscriptaddress,showpacs]{revtex4}

\usepackage{graphicx, latexsym, verbatim}
\usepackage{amssymb, amsmath}
\usepackage[ansinew]{inputenc}
\usepackage{color}

\newcommand{\ud}{\mathrm{d}}

\newcommand{\Hb}{\mathbf{H}}
\newcommand{\T}{\mathcal{T}}
\newcommand{\Kb}{\mathbf{K}}
\newcommand{\kb}{\mathbf{k}}
\newcommand{\Mb}{\mathbf{M}}
\newcommand{\Gb}{\mathbf{G}}

\newcommand{\Ib}{\mathbf{I}}

\newcommand{\bEqa}{\begin{eqnarray}}
\newcommand{\eEqa}{\end{eqnarray}}
\newcommand{\Tr}{{\rm Tr}}

\newcommand{\Sig}{\mathbf{\Sigma}}
\newcommand{\Gam}{\mathbf{\Gamma}}

\begin{document}

% Use the \preprint command to place your local institutional report
% number in the upper righthand corner of the title page in preprint mode.
% Multiple \preprint commands are allowed.
% Use the 'preprintnumbers' class option to override journal defaults
% to display numbers if necessary
%\preprint{0.1}

%Title of paper
%\title{Modeling thermoelectric properties in surface disordered silicon nanowires}
\title{Electron- and phonon transport in silicon nanowires: an atomistic approach to thermoelectric properties}

% repeat the \author .. \affiliation  etc. as needed
% \email, \thanks, \homepage, \altaffiliation all apply to the current
% author. Explanatory text should go in the []'s, actual e-mail
% address or url should go in the {}'s for \email and \homepage.
% Please use the appropriate macro foreach each type of information

% \affiliation command applies to all authors since the last
% \affiliation command. The \affiliation command should follow the
% other information
% \affiliation can be followed by \email, \homepage, \thanks as well.

\author{Troels Markussen}
%\email[]{tma@mic.dtu.dk}
%\homepage[]{Your web page}
%\thanks{}
\affiliation{Department of Micro- and Nanotechnology, Technical University of Denmark, DTU Nanotech, Building 345 East, DK-2800 Kgs. Lyngby, Denmark}

\author{Antti-Pekka Jauho}
\affiliation{Department of Micro- and Nanotechnology, Technical University of Denmark, DTU Nanotech, Building 345 East, DK-2800 Kgs. Lyngby, Denmark}
\affiliation{Department of Applied Physics, Helsinki University of Technology, P.O. Box 1100, FIN-02015 HUT, Finland}

\author{Mads Brandbyge}
\affiliation{Department of Micro- and Nanotechnology, Technical University of Denmark, DTU Nanotech, Building 345 East, DK-2800 Kgs. Lyngby, Denmark}

%Collaboration name if desired (requires use of superscriptaddress
%option in \documentclass). \noaffiliation is required (may also be
%used with the \author command).
%\collaboration can be followed by \email, \homepage, \thanks as well.	
%\collaboration{}
%\noaffiliation

\date{\today}

\begin{abstract}
We compute both electron- and phonon transmissions in thin disordered silicon nanowires. Our atomistic approach is based on tight-binding and empirical potential descriptions of the electronic and phononic systems, respectively.
Surface disorder is modeled by including surface silicon vacancies. It is shown that the average phonon- and electron transmissions through long SiNWs containing many vacancies can be accurately estimated from the scattering properties of the isolated vacancies using a recently proposed averaging method [Phys. Rev. Lett. {\bf 99}, 076803 (2007)]. We apply this averaging method to surface disordered SiNWs in the diameter range $1-3$ nm to compute the thermoelectric figure of merit, $ZT$. It is found that the phonon transmission is affected more by the vacancies than the electronic transmission leading to an increased thermoelectric performance of disordered wires, in qualitative agreement with recent experiments. The largest $ZT>3$ is found in strongly disordered $\langle111\rangle$ oriented wires with a diameter of 2 nm.
\end{abstract}

% insert suggested PACS numbers in braces on next line
%
% 72.10.Fk Scattering by point defects, dislocations, surfaces, and other imperfections (including Kondo effect)
% 72.15.Lh Relaxation times and mean free paths
% 73.63.-b Electronic transport in nanoscale materials and structures (see also 73.23.-b Electronic transport in mesoscopic systems)
% (73.21.Hb Quantum wires)
% 63.22.-m Phonons or vibrational states in low-dimensional structures and nanoscale materials  
% 63.22.Gh Nanotubes and nanowires  
% 66.70.-f Nonelectronic thermal conduction and heat-pulse propagation in solids; thermal waves  

\pacs{63.22.Gh, 66.70.-f, 73.63.-b, 72.10.Fk}
% insert suggested keywords - APS authors don't need to do this
%\keywords{}

%\maketitle must follow title, authors, abstract, \pacs, and \keywords
\maketitle

\section{Introduction}
Both electronic and thermal transport in quasi one-dimensional systems such as carbon nanotubes and semiconducting nanowires are of great importance for their applications. Often a large electronic conductance, $G_e$, and a large thermal conductance, $G_{ph}$ is desired in electronic components. However, for thermoelectric applications, the thermal conductance should be as small as possible to increase the efficiency. The performance of a thermoelectric device is characterized by the figure of merit, $ZT=S^2\sigma\,T/\kappa$, where $S$ is the Seebeck coefficient, $\sigma$ the electronic conductivity, $T$ the temperature, and $\kappa$ is the total thermal conductivity. A good thermoelectric device should have $ZT>3$~\cite{MajumdarScience2004}. The ideal thermoelectric material is a so-called 'phonon-glass-electron-crystal' (PGEC) with a high thermopower, $S^2\,\sigma$ and a low thermal conductivity, $\kappa$. Different approaches have been followed to increase ZT including superlattices, alloys, embedded nano-particles in both 3D, 2D, and 1D structures~\cite{DresselhausReview}. It was shown theoretically in Ref.~\cite{HicksDresselhaus1993} that the thermoelectric figure of merit is larger in one-dimensional structures than in bulk. Indeed, recent experiments showed very large figure of merits in silicon nanowires (SiNWs)~\cite{HochbaumNature2008,BoukaiNature2008} even though bulk silicon is a poor thermoelectric material. These experiments indicated that the phonon conductance was reduced more by surface disorder in SiNWs than the electronic conductance, leading to very high $ZT$. 

%In (quasi-) one-dimensional nanowires, surface scattering of both electrons and phonons become important for thin wires. Recent experiments 
From a theoretical point of view it is not obvious that surface roughness will affect the phonons more than the electrons, and a model that treats both effects on the same level of approximation is thus desired. In order to theoretically interpret experiments or possibly to design new thermoelectric materials or devices, it is thus important to treat both the electronic and the phononic transport on the same footing.  

Recently, Vo et al.~\cite{VoNanoLett2008} used \textit{ab initio} calculations and the Boltzmann equation to calculate $ZT$, and found values up to 8 in SiNW. The phonon heat conductance was, however, not calculated but used as a free parameter. However, we recently showed that the phonon heat conductance of pristine SiNWs is strongly anisotropic with $\langle110\rangle$ wires having up to two times larger conductance than $\langle100\rangle$ and $\langle111\rangle$ wires~\cite{MarkussenPristinePhononPaper}. We show in this article that the anisotropies remain in disordered wires, thus clearly effecting the thermoelectric properties of SiNWs. 

Phonon transport was recently modeled in nanowires with surface disorder \cite{MurphyPRB2007} and in SiNWs with an amorphous coating \cite{MingoYangPRB2003}. Also, several recent theoretical works have been concerned with scattering of electrons in SiNWs by surface roughness~\cite{SvizhenkoPRB2007,LherbierPRB2008} and defects or dopant impurities~\cite{BlasePRL,BlaseNanoLett2006,Markussen2006,MarkussenPRL2007}. We showed in Ref.~\cite{MarkussenPRL2007} that the average electronic conductance of an ensemble of \textit{long} SiNWs containing many randomly placed dopants could be accurately estimated from the scattering properties of the isolated dopants. This enables much faster calculations and the ability to study a larger diameter range. 

In this paper we demonstrate that the same single-defect averaging can be applied to both phonon- and electron transport in the case of surface vacancy scattering. The average thermal conductance of long SiNWs containing many vacancies can thus be accurately estimated from the scattering properties of the isolated vacancies. We use this knowledge to compute both the electronic- and phononic conductances in ultra-thin SiNWs with surface disorder and to calculate the thermoelectric figure of merit in SiNWs with diameters ranging from 1--3 nm. We apply an empirical potential model to describe the phonons while the electronic system is modeled by a nearest neighbour tight-binding Hamiltonian. Both electron- and phonon conductances are calculated within the non-equilibrium Green's function (NEGF) formalism \cite{Haug08}.

Our microscopic theory confirms the experimental trend~\cite{HochbaumNature2008}: the electrons are less affected by surface disorder than the phonons, and the thermoelectric performance increases for increasing disorder (increasing number of vacancies). We consider wires in the $\langle100\rangle$, $\langle110\rangle$, and $\langle111\rangle$ directions and find that $\langle111\rangle$ oriented wires have the highest thermoelectric figure of merit with an optimum diameter at $2.0\,$nm. To simplify our analysis, we neglect both phonon-phonon, electron-phonon and electron-electron scattering mechanisms. Due to the relative simplicity of our model, our results do not allow for a quantitative comparison with experiments. However, we believe that our results do shed light on both electron- and phonon surface scattering, and on the relative magnitudes of the two.

The paper is organized as follows: In Sect.~\ref{Method} we present the phonon empirical potential model and the electronic tight-binding model. Also, we explain how to calculate the electronic- and phononic transmissions through long wires either by a recursive method or, simply from the transmission through single, isolated vacancies. %Finally we present formulas for the thermoelectric figure of merit, $ZT$.
In Sect. \ref{results} we present the results and end up by discussion and conclusions in Sect. \ref{conclusion}.

\section{Method} \label{Method}
As a model of surface disorder we introduce surface silicon vacancies, i.e. removed surface silicon atoms. 
This is a very simple model and deviates from the standard description of surface roughness, where the thickness of the wire diameter fluctuates on some characteristic length scale along the wire \cite{LherbierPRB2008}.
%This is a very simple model of surface disorder and different from surface roughness, where the wire diameter fluctuates along the wire. 
However, the simplicity of the vacancy disorder allows us to study single vacancy scattering and apply the averaging methods of Ref.~\cite{MarkussenPRL2007} and in this way study a larger diameter range than would be computationally manageable with our present methods and implementations.

\subsection{Phonon empirical potential model} \label{PhononModel}
The phonon system is described by an atomistic model with the interatomic potentials parameterized by the Tersoff empirical potential (TEP) model \cite{Tersoff1988} as implemented in the "General Lattice Utility Program" (\textsc{gulp})\cite{Gulp}. We use \textsc{gulp} to relax the atomic structure and to output the dynamical matrix, $\Kb$, for the relaxed system. Since the Tersoff potential is limited to nearest neighbour interactions, $\Kb$ can be written in a block-tridiagonal form:
\begin{eqnarray}
\Kb = \left(
\begin{array}{c c c c}
\kb_{11} & \kb_{12} & 0 & 0 \\
\kb_{21} & \kb_{22} & \kb_{23} & 0 \\
0 & \kb_{32} & \kb_{33} & \kb_{34} \\
0 & 0 & \kb_{43} & \kb_{44}
\end{array}
\right), \label{tridiagform}
\end{eqnarray}
where the sub-matrices $\kb_{ii}$ and $\kb_{i,i\pm1}$ describe the force constants within a unit cell and between neighbouring unit cells, respectively. The calculations use periodic boundary conditions and the super cell method. Neighbouring, parallel wires are sufficiently separated~\cite{separation} and therefore do not interact. A pristine, defect free wire can be modeled using only a single unit cell in the super cell. We shall denote by $\kb_{00}$ and $\kb_{01}$ the dynamical matrices describing the pristine wire unit cell and the coupling between neighbouring unit cells, respectively. Below, Fig. \ref{bandstructures} illustrates the unit cells.

When modeling vacancies, larger super cells are needed in order to avoid interactions between the periodically repeated vacancies. We find that 5 unit cells in the super cell~\cite{5unitCellComent} is sufficient for the scattering properties to be converged (see also Fig. \ref{longWireSetup}). We will denote the dynamical matrix around a given vacancy at position $\nu$ by $\widetilde{\Kb}^\nu$. When calculating the phonon properties we model pure silicon wires without any surface passivation. We have recently shown that including hydrogen on the surface only leads to insignificant changes~\cite{MarkussenPristinePhononPaper}. Also, we showed in Ref.~\cite{MarkussenPristinePhononPaper} that the thermal conductance of pristine SiNWs obtained with the TEP model agreed quantitatively with the more elaborate density functional theory (DFT) calculations.

\subsection{Electronic tight-binding model} \label{TBmodel}
The electronic system is described by a nearest neighbour $sp^3d^5s^*$ tight-binding (TB) basis. Contrary to the phonon system, it is necessary to include surface passivation in the electronic description in order to saturate the Si dangling bonds, and to avoid localized states inside the band gap. Often hydrogen is used for passivation in DFT based calculations and we adopt this strategy by including a hydrogen-like atom with a single $s$--orbital. The purpose of the hydrogen is only to passivate the dangling bonds, and not to describe the Si--H interactions in detail. The Si--Si TB parameters are from Ref.~\cite{Boykin2004} while the Si--H interactions are the same as in Ref.~\cite{Lake2005}. The same Si-Si tight-binding parameters were recently applied to model surface roughness in SiNWs transistors \cite{LuisierAPL2007}.
Due to the nearest neighbour model, the electronic Hamiltonian, $\Hb$, has a block tridiagonal form as in Eq.~\eqref{tridiagform}, in analogy with the dynamical matrix, $\Kb$. When a Si surface vacancy is introduced, we passivate the resulting Si dangling bonds with extra hydrogen atoms.

\subsection{Electronic conductance} \label{Electronic_Conductance}
We calculate the electronic current using the non-equilibrium Green's function (NEGF) formalism \cite{Haug08Ch12}. As usual in NEGF we divide our system into a left, central, and right region. The left and right contacts are modeled as two semi-infinite, defect free wires, and can be taken into account via the self-energies $\Sig_{L,R}^r(E)$~\cite{Markussen2006}, and the retarded Green's function in the central region is calculated as
\begin{equation}
\Gb^r(E) = [\,E\,\Ib-\Hb_C-\Sig_L^r(E)-\Sig_R^r(E)]^{-1}, \label{Gret}
\end{equation}
where $\Hb_C$ is the Hamiltonian in the central region containing defects, and $\Ib$ is the identity matrix. The transmission function through the central region is given by
\begin{equation}
\mathcal{T}(E) = \Tr[\,\Gam_L(E)\,\Gb^r(E)\,\Gam_R(E)\,\Gb^a(E)], \label{TransmissionEq}
\end{equation}
where $\Gam_{L,R}(E) = i(\Sig_{L,R}^r(E)-\Sig_{L,R}^a(E))$. The electronic current, $I$ is given by the Landauer formula,
\begin{equation}
I = \frac{2e}{h}\int_{-\infty}^\infty \ud E\,\mathcal{T}(E)\left[f(E,\mu_L)-f(E,\mu_R)\right] \label{I-formula},
\end{equation}
where $f(E,\mu_{L,R}) = 1/\left(\exp\left[(E-\mu_{L,R})/k_BT\right]+1\right)$ is the Fermi-Dirac distribution function in the left and right leads with chemical potentials $\mu_{L,R}$. The voltage difference between left and right is $V=(\mu_L-\mu_R)/e$. The linear electronic conductance, $G_e(\mu) = dI/dV$, is written as
\begin{equation}
G_e(\mu) = \frac{2e^2}{h}\int_{-\infty}^\infty \ud E\,\mathcal{T}(E)\left(-\frac{\partial f(E,\mu)}{\partial E}\right)=e^2L_0 \label{Ge},
\end{equation}
where we have introduced the function $L_m(\mu)$ to be used later:
\begin{equation}
L_m(\mu) = \frac{2}{h}\int_{-\infty}^\infty \ud E\,\mathcal{T}(E)(E-\mu)^m\left(-\frac{\partial f(E,\mu)}{\partial E}\right) \label{Lm}.
\end{equation}

\subsection{Phononic thermal conductance} \label{Phonon_conductance}
The phonon transmission can be calculated in a mathematically similar way as the electronic transmission using the substitutions
\begin{eqnarray}
E\,\Ib &\rightarrow& \omega^2\,\Mb  \\
\Hb_C &\rightarrow& \Kb_C
\end{eqnarray}
in Eq.~\eqref{Gret}. Here $\Mb$ is a diagonal matrix with elements corresponding to the masses of the atoms and $\Kb_C$ is the dynamical matrix of the central region. The left and right contacts are again modeled as two semi-infinite wires with self-energies $\Sig_{L,R}(\omega)$~\cite{WangPRE2007}. Note that we use the same symbols for electron- and phonon Green's functions and self-energies, but with different arguments: $\Gb(E)$ for electrons and $\Gb(\omega)$ for phonons. We stress that using \eqref{TransmissionEq} for is only valid when anharmonic (phonon-phonon) scattering can be neglected. The electronic analogue, Eq.~\eqref{TransmissionEq}, is likewise limited to mean-field theories \cite{Haug08Ch12}. In bulk Si, the room temperature anharmonic phonon-phonon relaxation length at the highest frequencies is $\lambda_a(\omega_{max})\sim 20\,$nm and increases as $\lambda_a\propto\omega^{-2}$ at lower frequencies \cite{MingoYangPRB2003}. For relatively short disordered wires, the anharmonicity effect is thus of limited importance.

The phonon thermal current at temperature $T$ can be calculated from the transmission function as \cite{WangPRE2007,YamamotoPRL2006,MingoPRB2006}
\begin{equation}
J_{ph}(T) = \frac{\hbar}{2\pi}\int_{0}^\infty\ud\omega\,\omega\,\mathcal{T}(\omega)\,[n_B(T_L)-n_B(T_R)],
\end{equation}
where $n_B(T) = (e^{\hbar\omega/k_BT}-1)^{-1}$ is the Bose-Einstein distribution and $T_{L,R} = T \pm \Delta T/2$.
In the limit of small temperature difference $\Delta T$, the phonon thermal conductance $\kappa_{ph}(T) = \frac{J_{th}}{\Delta T}$ is
\begin{equation}
\kappa_{ph}(T) = \frac{\hbar^2}{2\pi k_B T^2}\int_{0}^\infty\ud\omega\,\omega^2\,\mathcal{T}(\omega)\,\frac{e^{\hbar\omega/k_BT}}{(e^{\hbar\omega/k_BT}-1)^2}. \label{ThermalConductance}
\end{equation}
%For a pristine wire with perfect transmission, in the low temperature limit we can approximate $\mathcal{T}(\omega)=\mathcal{T}(0)=4$, since there are four acoustic modes with $\hbar\omega\rightarrow 0$ in the long wavelength limit \cite{ClelandBook}. The remaining integral in \eqref{ThermalConductance} can be calculated analytically to yield the universal thermal conductance quantum $G_Q=4(k_B^2\pi\, T/6\hbar)$.

%Finally, we can substitute the conductivities in Eq. \eqref{ZTdef} with the conductances \eqref{Ge} and \eqref{ThermalConductance} to get
%\begin{equation}
%ZT = \frac{S^2\,G_e\,T}{G_{ph} + \frac{\pi^2\,k_B^2\,T}{3e^2}G_e }.
%\end{equation}

\subsection{Modeling long wires by recursion}  \label{Recursion}
Both the electron- and phonon transmissions through longer wires containing more vacancies at random positions can be calculated using the recursive Green's function method (RGF), as described in \cite{Markussen2006}. By repeatedly adding either pristine wire parts or parts from different vacancy calculations, a long wire is "grown". Ensemble averaged properties are found by repeating the calculations for many wires with different configurations of the vacancy positions. Typically, several hundred calculations are needed to converge the sample averaged properties, which makes the long wire calculations very time consuming.

\begin{figure}[htb!]
\includegraphics[width=.43\textwidth]{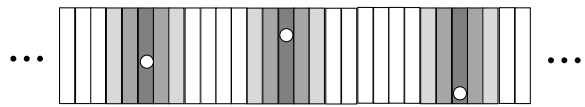}
    \caption{Sketch of a wire segment containing three different vacancies, represented by white circles. Each box represents a wire unit cell, and the two neighboring unit cells at each side of a vacancy are slightly modified as compared to the pristine wire unit cell, represented by white boxes.}
\label{longWireSetup}
\end{figure}

A schematic illustration of a segment of a long wire with randomly placed vacancies is shown in Fig. \ref{longWireSetup}. Each small box represents a wire unit cell and the three white circles represent three differently placed vacancies. As indicated by the shading, the two neighboring unit cells at each side of a vacancy are slightly modified as compared to the pristine wire unit cell, represented by white boxes.
The dynamical matrix (or Hamiltonian) inside each box is one of the diagonal sub-matrices in \eqref{tridiagform}, $\kb_{ii}$ and the coupling between two neighbouring boxes is described by the off-diagonal matrices, $\kb_{i,i+1}$. We emphasize that when we model wires with more than one vacancy, we assume that the individual vacancies are separated in space and do not interact. This limits the minimum vacancy-vacancy distance to $~2\,$nm in the axial direction. This method has recently been applied to calculate the electronic transmissions through long SiNWs~\cite{Markussen2006,MarkussenPRL2007,MarkussenJCompElec2008}.

\subsection{Modeling long wires from single defects} \label{Long_singles}
Recently we showed that the length dependent sample-averaged electronic transmission, $\bar{\T}$, of SiNWs with randomly placed dopant atoms ('defects') could be accurately estimated from the transmissions through the single defects as~\cite{MarkussenPRL2007,MarkussenJCompElec2008}:
\begin{equation}
\bar{\T} = \frac{\langle \T\rangle}{N_{\rm def}+(1-N_{\rm def})\frac{\langle \T \rangle}{\T_0}}, \label{Tsingle}
\end{equation}
where $N_{\rm def}=nL$ is the total number of defects in a wire of length $L$ with defect density $n$. $\T_0$ is the ideal ballistic transmission of the pristine wire, and $\langle \T\rangle = (\sum_{\nu=1}^M\T_\nu)/M$ is the average transmission of the $M$ different, isolated vacancies. Note that all the transmissions are energy (frequency)  dependent. In the short wire limit, $N_{\rm def}\rightarrow 0$ the transmission equals that of a pristine wire $\bar{\T}=\T_0$. For $N_{\rm def}=1$, $\bar{\T}=\langle \T\rangle$, while in the diffusive regime $N_{\rm def}\gg 1$, $\bar{\T}\propto 1/L$. The use of Eq. \eqref{Tsingle} is limited to the quasi-ballistic and diffusive regimes, where the inverse transmission, $1/\mathcal{T}$ increases linearly with wire length.

%Using the estimate \eqref{Tsingle} instead of sample averaging, the computation time is reduced by $\sim10^3$ for 100 nm long wires \cite{speedUpExplanation}.

\begin{figure}[htb!]	 
	\includegraphics[width=\columnwidth]{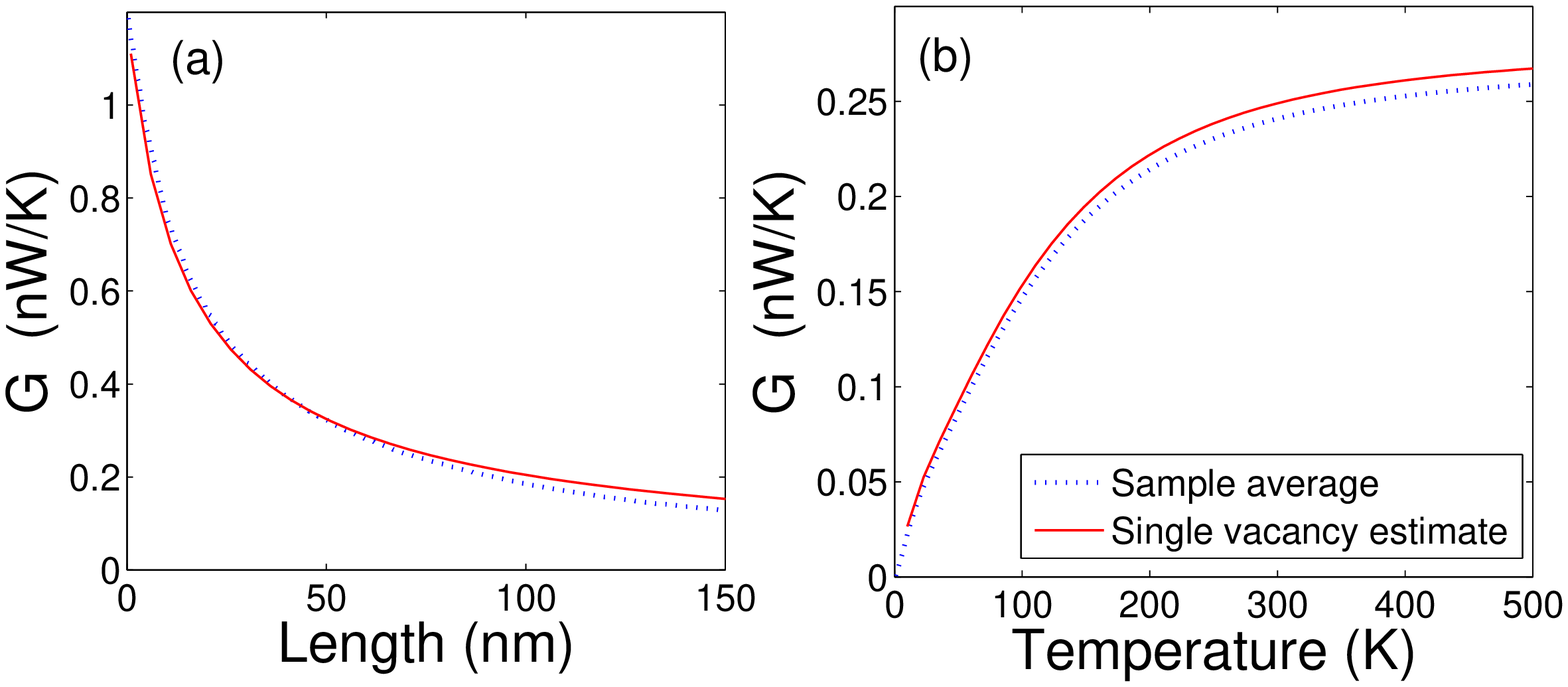}
	\includegraphics[width=\columnwidth]{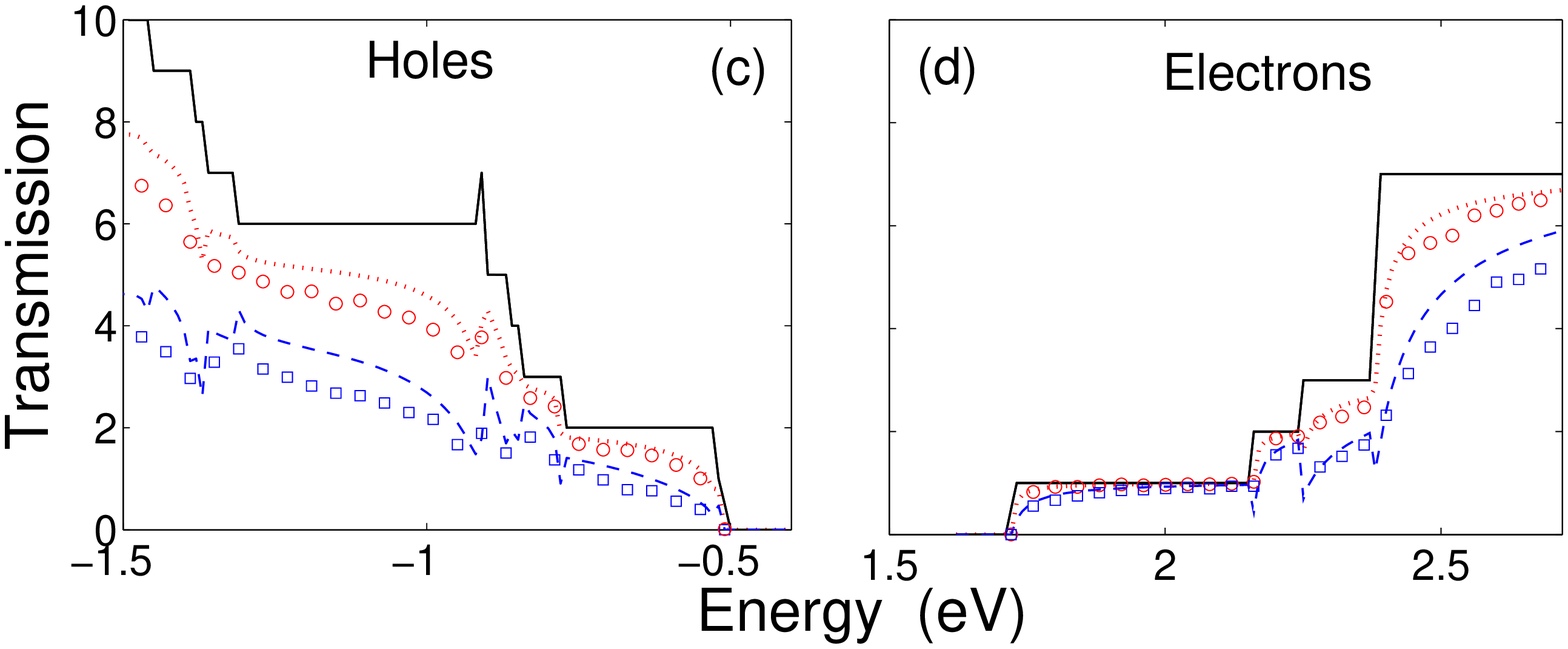} 	
	\caption{(color online). Thermal conductance and electronic transmission for 1.2 nm diameter $\langle110\rangle$ wires. Top row: Thermal conductance vs. length (a),  at $T=300\,$K and vs. temperature (b) at $L=75\,$nm. Dashed line: Long wire calculations and sample averaging. Solid line: results using Eq. \eqref{Tsingle}. Bottom row: Electronic transmission vs. energy for holes (c) and electrons (d). Solid line: pristine wire transmission; single vacancy estimates with $N_{\rm vac}=5$ (dotted red) and $N_{\rm vac}=20$ (dashed blue). The markers show sample average results for $N_{\rm vac}=5$ (red circles) and $N_{\rm vac}=20$ (blue squares).}
\label{singleVsAverage}
\end{figure}

Figure \ref{singleVsAverage} shows the thermal conductance vs. length (a) and vs. temperature (b). The results are obtained for a 1.2~nm diameter $\langle110\rangle$ wire with a mean vacancy-vacancy separation of 3.8~nm. The dashed line is obtained by sample-averaging the transmission of 200 wires, each 150~nm long (400 unit cells) with different vacancy positions, while the solid lines are obtained from Eq. \eqref{Tsingle}. It is evident that the sample-averaged conductance can be accurately estimated from the single-vacancy calculations.  In this specific case, the number of matrix inversions required for the sample average calculations is $>10^3$ times larger than the single vacancy-estimates.

In Fig. \ref{singleVsAverage} (c) and (d) we show the sample averaged electronic transmission for wires with $N_{\rm vac}=5$ (red circles) and $N_{\rm vac}=20$ (blue squares) number of vacancies. The dotted red and dashed blue lines are obtained using Eq. \eqref{Tsingle} with $N_{\rm vac}=5$ and $N_{\rm vac}=20$, respectively (substitute $N_{\rm def}$ with $N_{\rm vac}$ in Eq. \eqref{Tsingle}). Again we see that the simple and fast single-vacancy method, Eq. \eqref{Tsingle}, provides a reasonable estimate of the time consuming sample averaging results. The rest of the results in this paper are obtained from the electron and phonon transmissions through single, isolated vacancies using \eqref{Tsingle} to calculate the length dependent transmission.
\subsection{Thermoelectric figure of merit}  \label{ZT}
The efficiency of a thermoelectric material can be characterized by the dimensionless figure of merit, $ZT$, given by \cite{Marder}
\begin{equation}
ZT = \frac{S^2G_e\,T}{\kappa_{ph}+\kappa_{e}} \label{ZTdef},
\end{equation}
where $S$ is the Seebeck coefficient, $G_e$ the electronic conductance, $T$ the temperature, $\kappa_{ph}$ and $\kappa_e$ are the phonon- and electron contributions to the thermal conductance, respectively. $G_e$ and $\kappa_{ph}$ are given by Eq. \eqref{Ge} and \eqref{ThermalConductance}, while $S$ and $\kappa_e$ can be calculated when the electronic transmission $\mathcal{T}(E)$ is known \cite{SivanImry1986,EsfarjaniPRB2006,LundeFlensberg2005}:
\begin{eqnarray}
S(\mu,T) &=&  \frac{1}{eT}\frac{L_1(\mu)}{L_0(\mu)} \label{S-formula} \\
\kappa_{e}(\mu) &=&  \frac{1}{T}\left( L_2(\mu)-\frac{(L_1(\mu))^2}{L_0(\mu)} \right)  \label{kappa_e},
\end{eqnarray}
where $L_m(\mu)$ is given by Eq. \eqref{Lm}. The phonon contribution to  $S$ caused by phonon drag, is ignored in the present work, as our focus is on surface scattering, and inclusion of phonon drag is significantly more demanding involving electron-phonon coupling. Recent experiments indicated that a large phonon drag could be responsible for large $ZT$ in SiNWs \cite{BoukaiNature2008}, and our calculated $S$ might be too low.

\section{Results} \label{results}
\begin{figure}[htb!]
\includegraphics[width=\columnwidth]{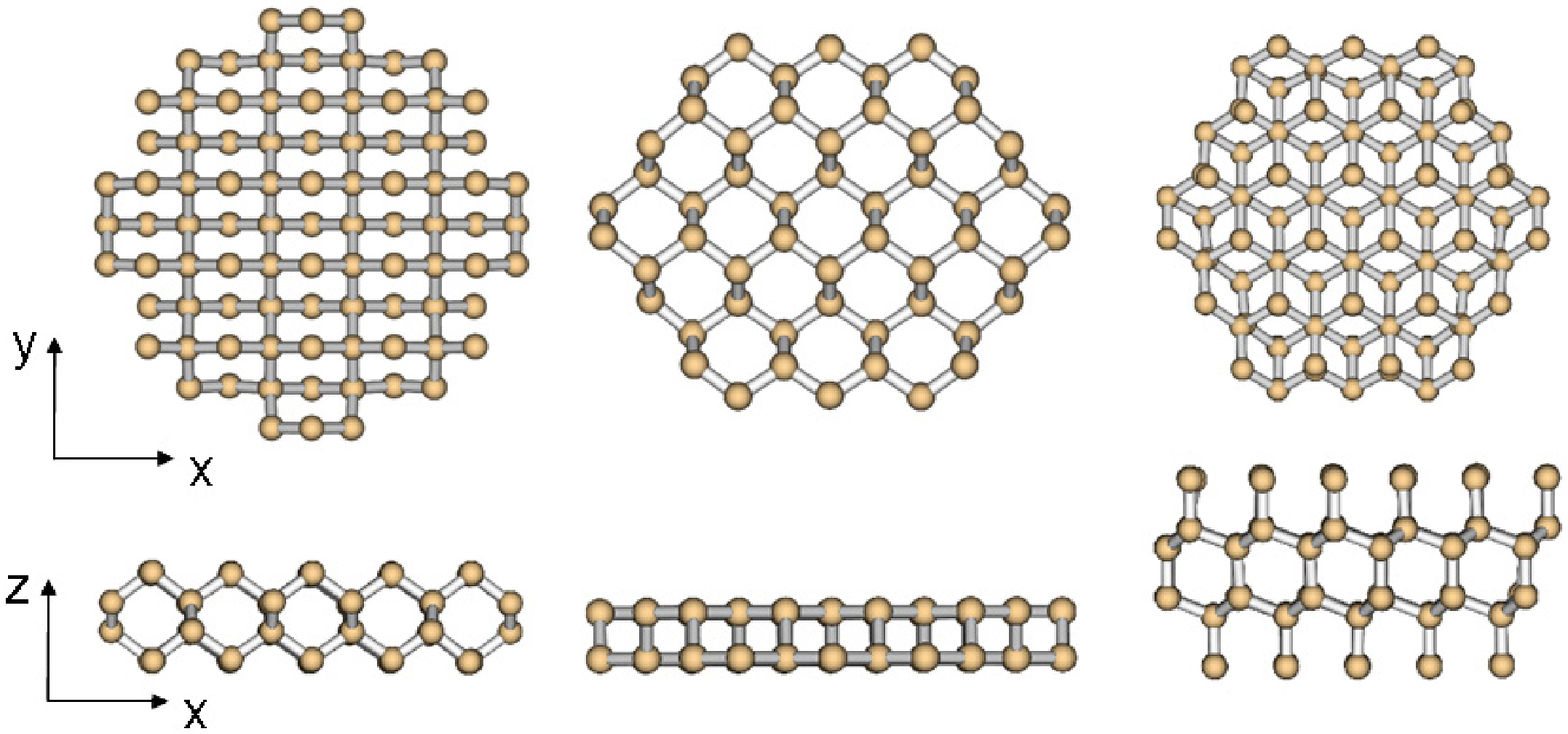}
\includegraphics[width=\columnwidth]{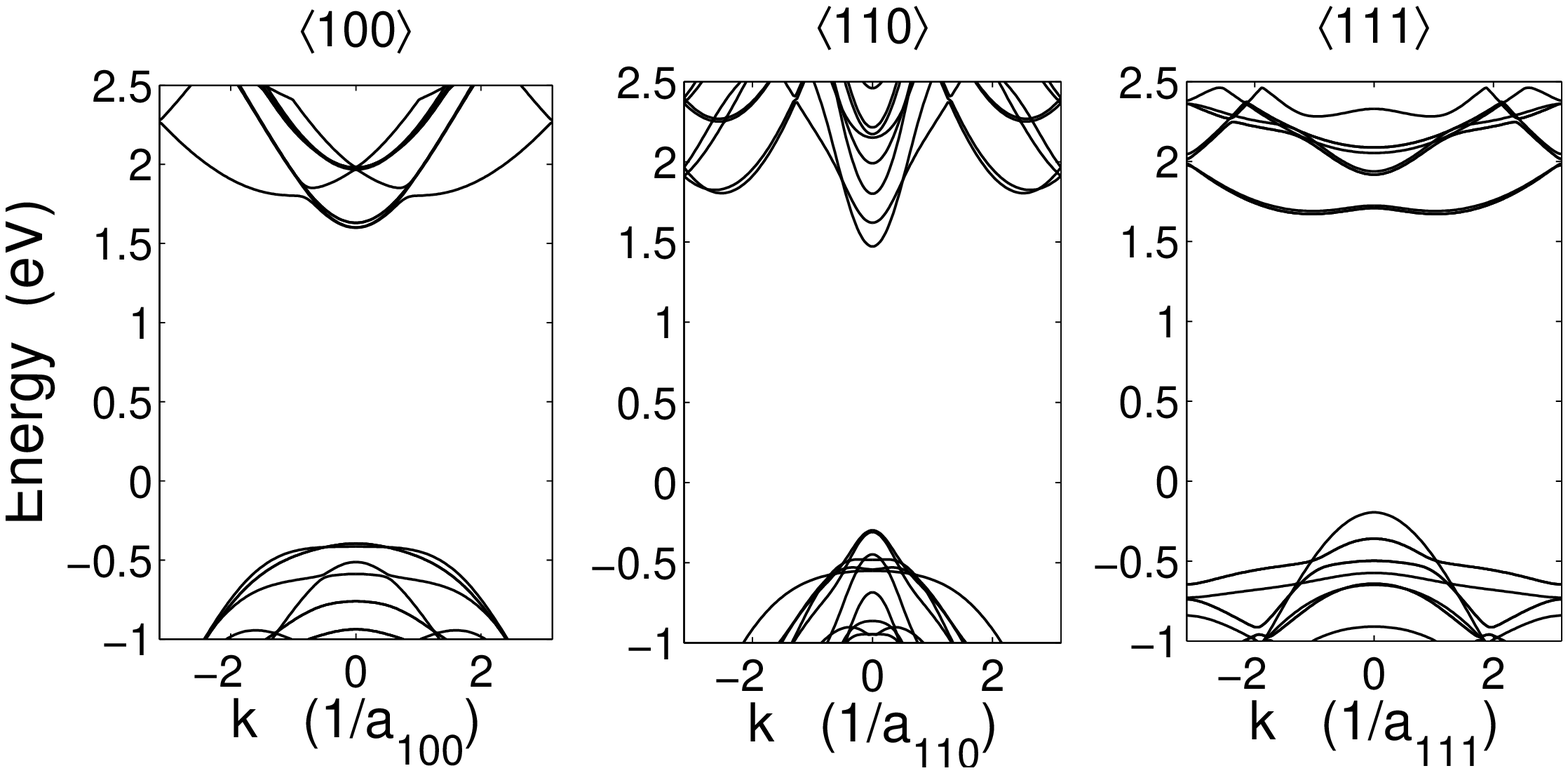}
    \caption{(color online). Cross sectional (top panel) and side views (middle panel) of  $\langle100\rangle$, $\langle110\rangle$, and $\langle111\rangle$ oriented wires (from left to right). The wires are oriented along the $z$-axis. Note that we have not plotted the passivating hydrogen atoms. Bottom panel: Electronic band structure around the band gap.}
\label{bandstructures}
\end{figure}

We consider SiNWs oriented along the $\langle100\rangle$, $\langle110\rangle$, and $\langle111\rangle$ directions. Cross sectional- and side views of the different wires are shown in Fig.~\ref{bandstructures} (top and middle panel). In the bottom panel we show the corresponding electronic band structure in the vicinity of the band gap. The corresponding phonon band structures can are reported in Ref.~\cite{MarkussenPristinePhononPaper}. In the following we show results for vacancy scattering in different wires. The transmissions calculated from Eq.~\eqref{Tsingle} are averaged over all inequivalent surface vacancy positions.

\subsection{Phonon- and electron vacancy scattering}
\begin{figure}[htb!]
\includegraphics[width=\columnwidth]{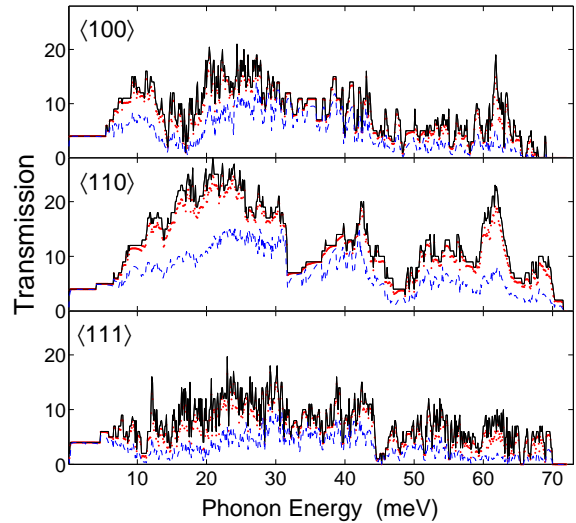}
    \caption{(color online). Average phonon transmission from Eq.~\eqref{Tsingle} through $D=2.0\,$nm wires containing 1 (dotted red) and 10 (dashed blue) vacancies. The solid black curve shows the pristine wire transmission. }
\label{Tph_vs_E}
\end{figure}

Figure \ref{Tph_vs_E} shows the average phonon transmission function through wires containing surface vacancies. The results are calculated from the single vacancies using Eq.~\eqref{Tsingle} with $N_{\rm def}=1$ (dotted red) and $N_{\rm def}=10$ (dashed blue). The solid black curve is the pristine wire ($N_{\rm def}=0$) transmission.
Observe that the four acoustic modes at $\hbar\omega\lesssim 4\,$meV are unaffected by the vacancies and transmit almost perfectly. %At other energies the phonons are scattered at different degrees, depending on the energy.
Generally, the scattering is relatively weak in the interval $30\lesssim \hbar\omega\lesssim40\,$meV, while a significantly stronger scattering is observed between 10-30 meV.
Notice also that the $\langle110\rangle$ wire has the largest transmission while the $\langle111\rangle$ wire has the smallest. This anisotropy was recently analyzed in Ref.~\cite{MarkussenPristinePhononPaper} for pristine wires. Based on the anisotropic phonon transmissions in Fig.~\ref{Tph_vs_E} we would expect $\langle111\rangle$ wires to have the highest $ZT$ values since they have the lowest thermal conductance. On the other hand, $\langle110\rangle$ wires are expected to have the lowest $ZT$ values since they have the highest thermal conductance.

\begin{figure}[htb!]
\includegraphics[width=\columnwidth]{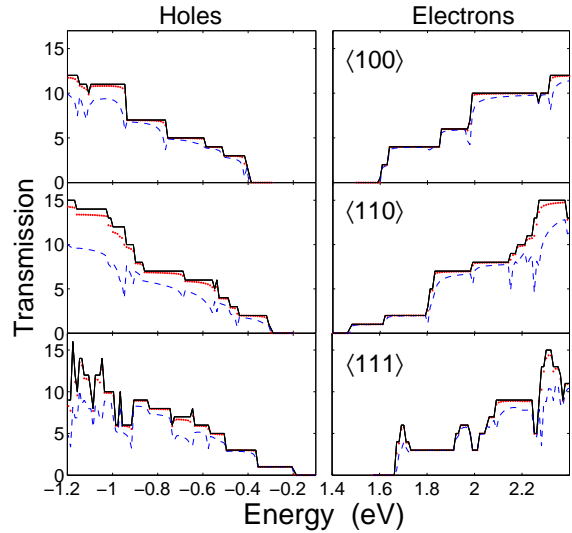}
    \caption{(color online). Average hole- and electron transmission from Eq.~\eqref{Tsingle} through $D=2.0\,$nm wires containing 1 (dotted red) and 10 (dashed blue) vacancies. The solid black curve shows the pristine wire transmission. }
\label{Te_vs_E}
\end{figure}

%\subsection{Electron vacancy scattering}
Figure \ref{Te_vs_E} shows the average electronic transmission from Eq.~\eqref{Tsingle}, again with $N_{\rm def}=1$ (dotted red) and $N_{\rm def}=10$ (dashed blue). The solid black curve shows the pristine wire transmission. Left and right panels correspond to hole- and electron transmissions in the valence- and conduction bands, respectively. Generally, the holes are scattered more than the electrons, and we would thus expect n-type wires to have higher $ZT$ values than p-type wires. %[More...???]

\subsection{$ZT$ calculations}
Figures \ref{Tph_vs_E} and \ref{Te_vs_E} indicate that both electrons and holes generally are less affected by the vacancies than the phonons, implying that increasing the surface disorder (increasing $N_{\rm def}$ either by increasing the vacancy density or the wire length) would increase the $ZT$ values.
\begin{figure}[htb!]
\includegraphics[width=.8\columnwidth]{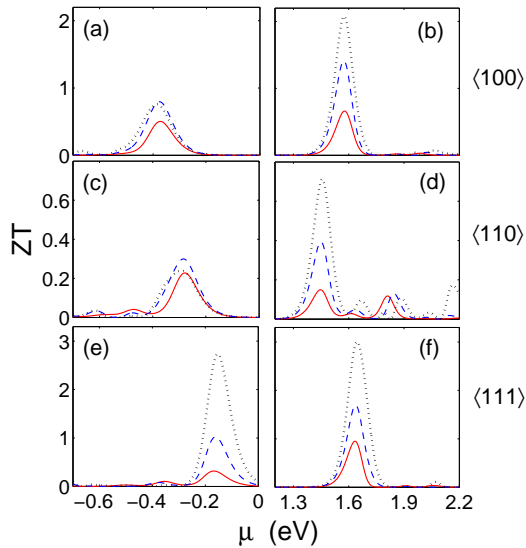}
    \caption{(color online). Calculated $ZT$ values vs. chemical potential at $T=300\,$K. Solid red:  $N_{\rm vac}=10$;  dashed blue: $N_{\rm vac}=100$; dotted black: $N_{\rm vac}=1000$.}
\label{ZTfig}
\end{figure}
This is indeed what we observe in Fig. \ref{ZTfig} showing calculated $ZT$ values as a function of chemical potential at different degrees of disorder. The temperature is $T=300\,$K. Both for electrons and holes in all wire directions, we observe an increase in $ZT$ as the number of vacancies is increased from 10 (solid red) to 100 (dashed blue). At $N_{\rm vac}=1000$ (dotted black) the $ZT$ is increased in all n-type wires (electrons) while only the $\langle111\rangle$ wire shows increased hole $ZT$. As expected, the $\langle111\rangle$ wires have the largest $ZT$ values, while the $\langle110\rangle$ wires have the smallest. This is partly due to the anisotropic phonon heat conductance, and partly due to difference in the electronic band structure. While both the $\langle100\rangle$ and $\langle110\rangle$ wires have the conduction band minimum (CBM) at $k=0$, the $\langle111\rangle$ wire has an indirect band gap, with the CBM at $k\neq 0$. This implies that close to the conduction band edge, there is a peak with six conducting channels in the $\langle111\rangle$ wire. The $\langle100\rangle$ wire quickly reaches a transmission plateau of four, whereas the $\langle110\rangle$ wire has a relatively broad plateau with only two channels.

The maximum $ZT$ occurs at a chemical potential close to the band edge, which requires a high doping concentration. The dopant atoms would certainly affect the electronic conductance~\cite{MarkussenPRL2007,MarkussenJCompElec2008,RuraliNanoLett2008} and, depending on the dopant type, also the phonon transmissions. Using phosphorous or aluminum as n- and p-type dopants would probably affect the phonons relatively little due to the small atomic mass difference between P/Al and Si, whereas nitrogen or boron are expected to scatter the phonons relatively more. As our main focus in on the surface disorder scattering the present calculations do not take the dopant scattering into account, and the computed values for $G_e$ give an upper limit.

\begin{figure}[htb!]
\includegraphics[width=.95\columnwidth]{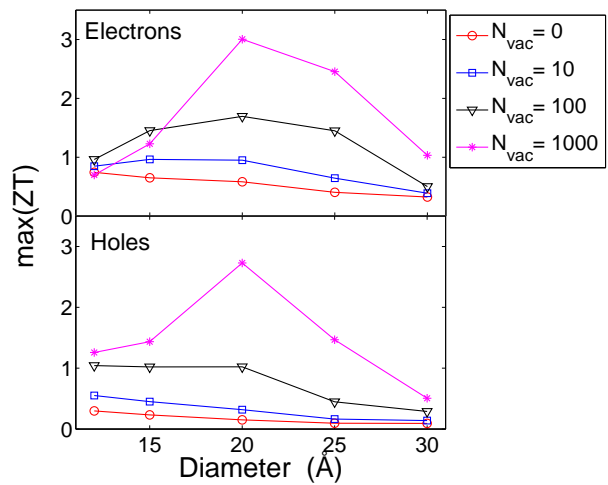}
    \caption{(color online). Diameter dependence of the maximum $ZT$ in $\langle111\rangle$ wires with different numbers of vacancies }
\label{ZT_vs_diam}
\end{figure}

Figure \ref{ZT_vs_diam} shows the maximum $ZT$ vs. wire diameter, $D$, for $\langle111\rangle$ wires with different degrees of disorder (number of vacancies). For both n-type (electrons) and p-type (holes) pristine wires with $N_{\rm vac}=0$, the maximum $ZT$ increases with decreasing diameter. The reason is that at $T=300\,$K effectively all phonon modes contribute to the phonon heat conductance, which therefore scales as in the continuum limit, $\kappa_{ph}\propto A$, where $A$ is the cross sectional area. The electronic conductance, $G_e$, is more or less constant as long as the distance between the electronic bands is larger than $k_BT$, which is the case for the diameter range considered. For larger wires, $G_e$ will also increase.

When vacancies are introduced, we observe an optimal diameter at 2 nm for n-type wires (electrons) independently of disorder strength \cite{diameterscaling}. For diameters $>$ 20 \AA, we also observe that increasing the disorder increases the maximum $ZT$, while for the smallest wires the largest $ZT$ is found for $N_{\rm vac}=100$. At $D=20\,$\AA, and $N_{\rm vac}=1000$, the maximum $ZT$ is six times larger than in the pristine wire with the same diameter.
In the p-type wires (holes) with $N_{\rm vac}=10$ and $N_{\rm vac}=100$ the maximum $ZT$ increases for decreasing wire diameter. For the strongest disorder ($N_{\rm vac}=1000$) we again see a maximum at $D=20\,$\AA, where $ZT$ is 18 times larger than in the pristine wire.

\section{Discussion and Conclusion} \label{conclusion}
\begin{figure}[htb!]
\includegraphics[width=.75\columnwidth]{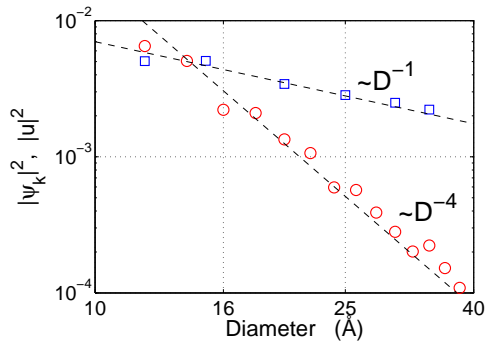}
    \caption{(color online). Diameter dependence of the average weight of the lowest conduction band electron Bloch state, $|\psi_k(r_{surf})|^2$ (circles), and average of all phonon eigenmodes, $|\bar{u}(r_{surf})|^2$, (squares) at the surface atoms in $\langle111\rangle$ wires. Notice the log-log scale.}
\label{weights_vs_diam}
\end{figure}

It is interesting that an optimal diameter exists where $ZT$ is maximized. To understand this we note that the weight of electron Bloch states, $|\psi_k|^2$, and phonon eigenmodes, $|u|^2$, on the surface atoms have a different diameter dependence as shown in Fig. \ref{weights_vs_diam}. The weight of the lowest conduction band electronic Bloch states (in the $\Gamma$-point) on the surface atoms decays as $\sim D^{-4}$ (circles). We note that this is the same scaling as one obtains from the effective mass Schrödinger equation. The average amplitude of \textit{all} the phonon eigenmodes (also in the $\Gamma$-point) on the surface atoms scales, on the other hand, as $\sim D^{-1}$ (squares). 

In a Fermi's golden picture, the electron scattering rate $\gamma_e \propto|\psi_k(r_{vac})|^2$ of a vacancy at position $r_{vac}$, and likewise for the phonon scattering $\gamma_{ph}\propto|u(r_{vac})|^2$~\cite{MurphyPRB2007}. The different diameter scalings of $|\psi_k(r_{surf})|^2$ and $|\bar{u}(r_{surf})|^2$ imply that surface disorder affects the electrons relatively more for the thinner wires. This effect tends to \textit{decrease} $ZT$ for decreasing diameter. On the other hand, as discussed above, the phonon conductance scales as $\kappa_{ph}\propto D^2$ thus \textit{increasing} $ZT$ for decreasing diameter. Given the two opposing diameter dependencies, it is reasonable that an optimal diameter exists where $ZT$ is maximized.

Summarizing, we have used the averaging method proposed in Ref.~\cite{MarkussenPRL2007} and have shown that both the electron- and phonon transmission through long wires containing many randomly placed vacancies can be reproduced from the transmissions through isolated vacancies. This enables one to perform relatively fast calculations on wires with experimentally relevant lengths. The averaging method was in Ref.~\cite{MarkussenPRL2007} applied to scattering at dopant impurities, and with the present study of vacancies, we believe that its validity is general for any random arrangement of localized structural disorder (defects, impurities, vacancies, adatoms, physisorbed molecules, etc.) and for both electron- and phonon transmissions. It remains to be investigated whether a similar approach can be extented to systems where the wire diameter fluctuates along the wire.

SiNWs oriented along the $\langle111\rangle$ direction have the largest $ZT$ values while $\langle110\rangle$ wires have the smallest. This is primarily due to the anisotropic heat conductance in SiNWs~\cite{MarkussenPristinePhononPaper}. $\langle110\rangle$ wires have larger $\kappa_{ph}$ and thus a smaller $ZT$. Differences among the wire orientations in the electronic transmission and Seebeck coefficients do also play a role.

For the $\langle111\rangle$ wires, we observe an increasing maximum $ZT$ for decreasing the diameter down to $D=2.0\,$nm which indicates that decreasing the wire diameter from typical dimensions of 10-50 nm to the sub-10 nm range might increase the thermoelectric performance. More thorough analysis including more realistic surface roughness is however required before firm conclusions can be drawn. We will address this in a future communication. Also, a quantitative model should also include electron-phonon- as well as phonon-phonon scattering~\cite{MingoPRB2003}. However, we believe that the present results, where electron- and phonon-vacancy scattering is based on the same atomistic structure, is an important first step on the way.

\begin{acknowledgements}
We thank the Danish Center for Scientific Computing (DCSC) and Direkt\o r Henriksens Fond for providing
computer resources. TM acknowledge the Denmark-America foundation for financial support. APJ
is grateful to the FiDiPro program of the Finnish Academy.
\end{acknowledgements}

%\bibliography{bib1}

\begin{thebibliography}{35}
\bibitem[{\citenamefont{Troels}(2008{\natexlab{a}})}]{separation}
  \bibinfo{note}{In a typical calculation, the separation is 20 \AA.}
\bibitem[{\citenamefont{Troels}(2008{\natexlab{d}})}]{diameterscaling}
  \bibinfo{note}{When changing the wire diameter we keep the surface vacancy
  concentration constant. The number of vacancies listed in
  Fig.~\ref{ZT_vs_diam} correspond to $D=20\,$\AA. In general,
  $N_{vac}(D)=N_{vac}(20$\AA$)D/20$\AA.}
\end{thebibliography}

\end{document}